\documentclass[conference]{IEEEtran}
\IEEEoverridecommandlockouts
\usepackage{cite}
\usepackage{amsmath,amssymb,amsfonts}
\usepackage{graphicx}
\usepackage{textcomp}
\usepackage{xcolor}
\usepackage{tikz}
\usepackage{tcolorbox}
\usepackage{amsmath}
\usepackage{float}
\usepackage{algpseudocode}
\usepackage{algorithm}
\newcommand{\algrule}[1][.4pt]{\par\vskip.5\baselineskip\hrule height #1\par\vskip.5\baselineskip}
\usepackage{hyperref}

\hypersetup{
    colorlinks=true,
    linkcolor=black,
    citecolor=black,
    filecolor=black,
    urlcolor=black,
}

\usepackage[T1]{fontenc}
\usepackage{url}
\usetikzlibrary{shapes.multipart, arrows.meta, positioning}

\def\BibTeX{{\rm B\kern-.05em{\sc i\kern-.025em b}\kern-.08em
    T\kern-.1667em\lower.7ex\hbox{E}\kern-.125emX}}
\begin{document}

\title{Bisecting K-Means in RAG for Enhancing Question-Answering Tasks Performance in Telecommunications\\
}

\author{\IEEEauthorblockN{Pedro Sousa\IEEEauthorrefmark{1},
Cláudio Klautau Mello\IEEEauthorrefmark{1},
Frank B. Morte\IEEEauthorrefmark{1}, and
Luis F. Solis Navarro\IEEEauthorrefmark{2}}
\IEEEauthorblockA{\IEEEauthorrefmark{1}Federal University of Pará, 
Institute of Technology, Belém, Brazil}
\IEEEauthorblockA{\IEEEauthorrefmark{2}State University of Campinas, Institute of Computing, Campinas, Brazil}
\IEEEauthorblockA{\IEEEauthorrefmark{1}\{pedro.sousa, claudio.mello, frank.morte\}@itec.ufpa.br, \IEEEauthorrefmark{2}L214616@dac.unicamp.br }
}

% bloco agregado pra o pseudocode
\newcommand{\var}[1]{\text{\texttt{#1}}}
\newcommand{\func}[1]{\text{\textsl{#1}}}

\makeatletter
\newcounter{phase}[algorithm]
\newlength{\phaserulewidth}
\newcommand{\setphaserulewidth}{\setlength{\phaserulewidth}}
\newcommand{\phase}[1]{%
  \vspace{-1.25ex}
  % Top phase rule
  \Statex\leavevmode\llap{\rule{\dimexpr\labelwidth+\labelsep}{\phaserulewidth}}\rule{\linewidth}{\phaserulewidth}
  \Statex\strut\refstepcounter{phase}\textit{Pipeline~\thephase~--~#1}% Phase text
  % Bottom phase rule
  \vspace{-1.25ex}\Statex\leavevmode\llap{\rule{\dimexpr\labelwidth+\labelsep}{\phaserulewidth}}\rule{\linewidth}{\phaserulewidth}}
\makeatother

\setphaserulewidth{.7pt}

% fin do bloco pra pseudocode

% \author{\IEEEauthorblockN{Cláudio Klautau}
% \IEEEauthorblockA{\textit{Institute of Technology} \\
% \textit{Federal University of Pará}\\
% Belém, Brazil \\
% claudio.mello@itec.ufpa.br}
% \and
% \IEEEauthorblockN{Frank Bruno Ferreira Boa Morte}
% \IEEEauthorblockA{\textit{Institute of Technology} \\
% \textit{Federal University of Pará}\\
% Belém, Brazil  \\
% frank.morte@itec.ufpa.br}
% \and
% \IEEEauthorblockN{Luis Fernando Solis Navarro}
% \IEEEauthorblockA{\textit{Institute of Computing} \\
% \textit{State University of Campinas  }\\
% Campinas, Brazil \\
% l214616@dac.unicamp.br}
% \and
% \IEEEauthorblockN{Pedro Sousa}
% \IEEEauthorblockA{\textit{Institute of Technology} \\
% \textit{Federal University of Pará}\\
% Belém, Brazil \\
% pedro.sousa@itec.ufpa.br}

% }

\maketitle

\begin{abstract}
Question-answering tasks in the telecom domain are still reasonably unexplored in the literature, primarily due to the field's rapid changes and evolving standards. This work presents a novel Retrieval-Augmented Generation framework explicitly designed for the telecommunication domain, focusing on datasets composed of 3GPP documents. The framework introduces the use of the Bisecting K-Means clustering technique to organize the embedding vectors by contents, facilitating more efficient information retrieval. By leveraging this clustering technique, the system pre-selects a subset of clusters that are most similar to the user's query, enhancing the relevance of the retrieved information. Aiming for models with lower computational cost for inference, the framework was tested using Small Language Models, demonstrating improved performance with an accuracy of 66.12\% on phi-2 and 72.13\% on phi-3 fine-tuned models, and reduced training time.
\end{abstract}


\section{Introduction}

Large-scale language models (LLMs) have revolutionized many industries by providing better text understanding and generation capabilities. These models, such as Microsoft's phi-2 \cite{microsoft_phi2}, are trained on vast datasets with billions of parameters, allowing them to capture linguistic nuances and general knowledge in depth. LLMs applicability ranges from automating administrative tasks to personalized assistance, demonstrating their immense potential for innovation in many areas\cite{retrievalaugmentedgenerationlargelanguage}.

Despite their versatility, LLM models often need more knowledge in highly technical domains \cite{TelecomGPT}, such as telecommunications, due to the need for more specific data during their initial training. This can lead to inaccurate or inadequate answers when faced with questions that require deep, specialized understanding. To overcome these limitations, specialization techniques such as fine-tuning and Retrieval-Augmented Generation (RAG) are needed.

Fine-tuning involves retraining an existing LLM on a specialized dataset. This process adjusts the model’s weights to better align its predictions with domain-specific knowledge and terminology\cite{ovadia2023fine}. During fine-tuning, the model is repeatedly exposed to specialized data, allowing it to capture relevant patterns and nuances not present in the initial training. This method improves the model’s ability to provide accurate and appropriate answers to queries related to a specific domain.

RAG is another powerful technique used to increase the accuracy of LLMs. RAG combines text generation with information retrieval by integrating external data relevant to the query's context\cite{retrievalaugmentedgenerationlargelanguage}. This is accomplished through two main components: a retrieval model that searches for relevant documents in an external database and a generative model that uses these documents to produce contextualized answers\cite{lewis2021retrievalaugmented}. This approach expands the model's knowledge base and ensures that the answers are more accurate and informed by the latest, domain-specific context.

In the telecommunications sector, the application of specialized LLMs is particularly relevant \cite{bariah2023large}. This sector is characterized by its complexity and the need for specific terminology. A specialized LLM in this field can significantly reduce the time required to access crucial information, increase the accuracy of answers, and ensure compliance with international standards. This results in more efficient operations and the ability to quickly meet the technical demands of the sector.

This paper presents an alternative approach to specializing LLM models for the telecommunications sector. It uses the TeleQnA dataset \cite{TeleQnA}, created explicitly to assess the telecommunications knowledge of LLMs. The dataset consists of 10,000 questions and answers extracted from various sources, including standards specifications and research papers, ensuring broad coverage of the required knowledge. The goal is to develop a model capable of answering questions about 3GPP (The 3rd Generation Partnership Project) standards in release 18.

The proposal resides in integrating fine-tuning with clustering techniques. Specifically, Bisecting K-Means \cite{steinbach2000comparison}, to further refine the accuracy and relevance of the model. Bisecting K-Means is used to group similar data, identifying patterns and trends, that are critical for interpretation and analysis in telecommunications. The experimental results demonstrate significant improvements, highlighting the effectiveness of this methodology in adapting LLMs to highly specialized domains.

\begin{figure*}[ht]
    \centering
    \includegraphics[width=0.95\textwidth]{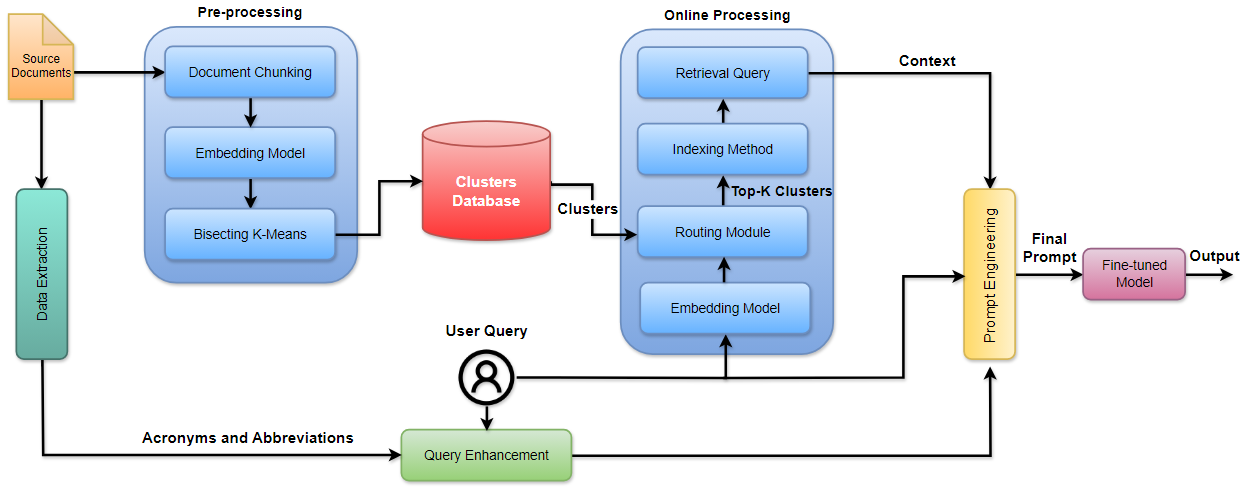}
    \caption{Diagram of the RAG pipeline that integrates a routing module with clustering techniques to efficiently retrieve the most relevant documents.}
\end{figure*}

\section{Related Works}

Recent studies have explored the application of LLMs in the telecom domain across various datasets, showcasing their potential to enhance numerous aspects of the industry. However, many of these studies fall short of addressing the optimization of search processes within knowledge represented by embedding vectors. For instance, Telco-RAG \cite{TelcoRAG} introduces the concept of leveraging LLMs for telecom datasets in question-answering tasks in a wide variety of information datasets \cite{TeleQnA} but lacks clarity on how specific components of the pipeline, such as embedding-based retrieval, semantic search, and response generation, contribute to overall performance improvements. This gap highlights the need for more in-depth research that delves into optimizing the search and retrieval mechanisms within LLMs to enhance their capabilities in telecom applications. 

TelecomRAG\cite{TelecomRAG}, for instance, is significantly constrained by the limitations of its pipeline, failing to address critical performance issues prevalent in the telecom sector, such as the extensive use of abbreviations and acronyms. This oversight can lead to inefficiencies and inaccuracies when processing data of this domain. In contrast, this work is designed to handle user entries that may pose challenges for an LLM to analyze effectively. The proposed framework incorporates a text preprocessing step that augments the prompt, ensuring that text generation is more accurate and reliable. This preprocessing enhances the LLM's ability to understand and generate meaningful responses, thereby improving the overall performance of question-answering tasks.

Authors of \cite{Telecom_RAG_IEEE} optimized the internal parameters of the RAG pipeline, particularly in the retrieval of indexes, demonstrating significant efficiency improvements. However, it remains limited by its reliance on the original structure of the content, offering no alternative methods for searching the knowledge representation. In contrast,  this work delves deeper into the knowledge representation by utilizing clustering techniques on the content, and in this way grouping the embeddings by the chunks content, disregarding their initial structure. This approach allows for more efficient searches within the embedding space, enhancing the overall performance and effectiveness of the information retrieval process.

%Finally, this work also contributes by applying the developed framework in smaller LLMs, aiming to consume fewer resources.

\section{Methodology}

Several critical steps were implemented to specialize an LLM model in answering questions about Telecom domain specifications. The documents used in this work were in the TeleQnA dataset, a compilation of 3GPP series documents in question-answering format with multiple-choice answers.

A series of methods were used to harness better performance. As shown in Figure 1, the RAG pipeline begins with the set of 3GPP source documents $\mathcal{A}$, which undergo a data extraction process to collect the set of acronyms and abbreviations $\mathcal{B}$. These elements, combined with the user's query, pass through a query enhancement process that augments the user's input for better accuracy. The source documents are then converted into the set of chunks $\mathcal{C}$ with chunk length $\alpha$ and passed through an embedding model to produce the embedding matrix $\textbf{E}$, followed by the Bisecting K-Means step with number of clusters $\beta$ that organizes the information by content into the clusters set $\mathcal{D}$. The clustered data is stored in a vector database, which in this work was SQLite3 \cite{sqlite2020hipp}, as part of the pre-processing block, which is performed only once.

In the online processing phase, the set of acronyms and abbreviations $\mathcal{B}$ and the user query are combined to create the enhanced query EQ. Also, in this process, each new user query is converted into an embedding matrix $\textbf{Q}$ and is compared with the clusters set $\mathcal{D}$ in the vector database through a routing module. This module selects a subset of clusters $\mathcal{K}$ most related to the user's query, being $|\mathcal{K}| = \gamma$, and retrieves the most relevant chunks by using an indexing method - in this case, the FAISS model \cite{faiss} - to obtain the chunks indexes vector $\textbf{t}$. These chunks, which compose the context set $\mathcal{R}$ that has cardinality $|\mathcal{K}| = \delta$, along with the original user query and the enhanced version EQ, are combined as the final prompt and are provided as the whole context through a prompt engineering process to a fine-tuned LLM, which then generates an accurate and contextually relevant response.

Algorithm 1 shows the entire process, divided into two parts: the first, which is processed only once to create the clusters, and the second, which is processed with each new user query.
\begin{algorithm}

  \caption{Question-Answering Algorithm}
  \begin{algorithmic}[1]
    \Statex{\hspace{-0.7cm} \textbf{Pre-processing Block}}
    \Require{$\mathcal{A}, \alpha, \beta$}
        \Procedure{Preprocessing}{$\mathcal{A}$}
        \State $\mathcal{B} \gets \text{DataExtraction}(\mathcal{A})$
        \State $\mathcal{C} \gets \text{DocumentChunking}(\mathcal{A}, \alpha)$

        \State $\mathbf{E} \gets \text{EmbeddingModel}(\mathcal{C}$)
        \State $\mathcal{D} \gets \text{BisectingKMeans}(\mathbf{E}, \beta)$
        
        \For {each cluster $\mathcal{F} \in \mathcal{D}$}
            \For {each embedding $\mathbf{e} \in \mathcal{F}$}
                \State $\text{Store}(\mathbf{e}, \text{VectorDB})$
            \EndFor
        \EndFor
    \EndProcedure
   \algrule
    \Statex{\hspace{-0.7cm} \textbf{Online Processing Block}} 
    \Require{UserQuery, $\mathcal{B}, \mathcal{D}, \gamma, \delta$}
    \Procedure{OnlineProcessing}{\text{UserQuery}}
         \State $\text{EQ} \gets \text{QueryEnhancement} (\text{UserQuery}, \mathcal{B})$
        \State $\mathbf{Q} \gets \text{EmbeddingModel}(\text{UserQuery}$)
        \State $\mathcal{K} \gets \text{RoutingModule}(\mathcal{D},$ $\mathbf{Q}$)
        \State $\textbf{t} \gets \text{IndexingMethod}(\mathcal{K}, \gamma)$
        \State $\mathcal{R}  \gets \text{RetrieveContext}(\textbf{t}, \delta)$
        \State $\text{FinalPrompt} \gets \text{PromptEngineering}(\text{UserQuery}, \mathcal{R}, \text{EQ})$
        
        \State $\text{Output} \gets \text{LLM}
        (\text{FinalPrompt})$
        
        \Return $\text{Output}$
    \EndProcedure
  \end{algorithmic}
\end{algorithm}

\subsection{Query Enhancement}

%Caracteristica del lenguaje usado en Telecomunicaciones

In the field of telecommunications, many words, abbreviations, and terms can be challenging to understand \cite{TelcoRAG}. Language models, such as phi-2, are usually trained with general data, so they might not grasp the specific technical terms found in 3GPP standard documents.

On the 3GPP website, we can find valuable information to enhance queries with the meanings of these terms. These documents contain detailed and specialized information, making it essential to teach these terms to the language models to help them understand and use the information correctly.
\begin{figure}[ht]
\centering
\begin{tikzpicture}[
  every node/.style={font={\normalsize}},
  box/.style={rectangle, draw, rounded corners, align=left, text width=8cm, inner sep=5pt},
  arrow/.style={-{Latex[length=3mm]}, line width=1pt}
]
\node[box, fill=white] (query) {\textbf{[Initial Query]}: \\ What functionality does \textcolor{red!70}{\textbf{LLDP}} provide in the \textcolor{red!70}{\textbf{TSN}} \textcolor{blue!60}{\textbf{Transport Network}}?};
\node[box, fill=blue!20, below=0.3cm of query] (s1) {\textbf{[Definitions]}:\\ Transport Network: Network infrastructure that provides connectivity and bandwidth for customer services.};
\node[box, fill=red!20, below=0.3cm of s1] (s2) {\textbf{[Abbreviations]}\\ LLDP: Link Layer Discovery Protocol.  TSN: Time-Sensitive Networking };
\node[box, fill=green!20, below=0.3cm of s2] (s3) {\textbf{[Final Query]}: \\
What functionality does LLDP provide in the TSN Transport Network?

Transport Network: Network infrastructure that provides connectivity and bandwidth for customer services.

LLDP: Link Layer Discovery Protocol. 

TSN: Time-Sensitive Networking
};
\end{tikzpicture}
\caption{The process of query enhancement before retrieving information and passing it to the LLM for generating an answer.}
\label{fig:query_evolution}
\end{figure}

The process of enhancing a query involves extracting the definitions and abbreviations from the 3GPP documentation and incorporating these meanings into the query. To achieve this, we create two dictionaries: one for technical terms and one for abbreviations. The glossary of technical terms includes definitions of phrases such as "System Networking," while the glossary of abbreviations provides meanings for terms like "5G." These resources are utilized to expand the query and provide better context when retrieving information relevant to the questions. In Figure \ref{fig:query_evolution}, we present an example of the query enhancement process using the technique mentioned above. 

\subsection{Preprocessing of Document Texts}
\begin{figure}[ht]
    \centering
    \includegraphics[width=1\linewidth]{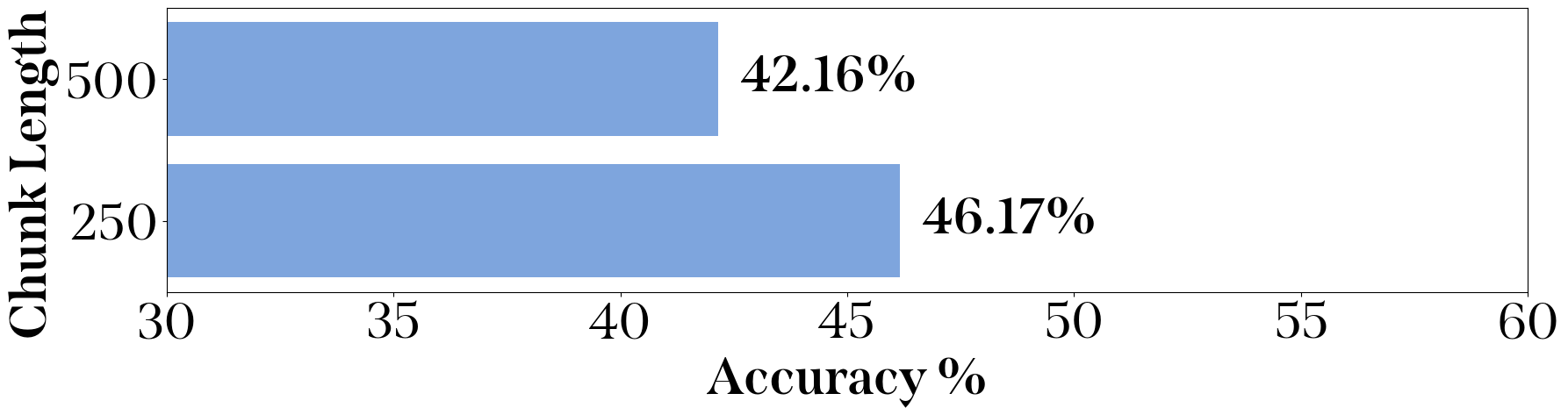}
    \caption{Performance of phi-2 on different chunk sizes.}
    \label{fig:Chunk_lenght}
\end{figure}

The texts from the 3GPP release 18 specifications were divided into chunks of various sizes, allowing a detailed and accurate analysis. Dividing texts into chunks is a common practice in natural language processing, as it facilitates the manipulation and analysis of large volumes of textual data\cite{yepes2024financialreportchunkingeffective}. It was chosen to test two chunk size configurations: 500 and 250 characters. These values were selected based on the results presented in \cite{TelcoRAG}, demonstrating that such division sizes effectively increase the chosen model's accuracy, as shown in Figure \ref{fig:Chunk_lenght}.

After dividing the texts into chunks, each segment was converted to embeddings, a crucial step for the language model to understand and use the data. Embeddings are vector representations of the texts, translating natural language into a form that machine learning models can process. This work uses the BAAI bge version large english model\cite{huggingfaceBGE} to map the chunks into numeric vectors, a model known for its efficiency in generating robust and accurate embeddings \cite{bge_embedding}.

\subsection{Clustering Methods for RAG}
\begin{figure}[ht]
    \centering
    \includegraphics[width=1\linewidth]{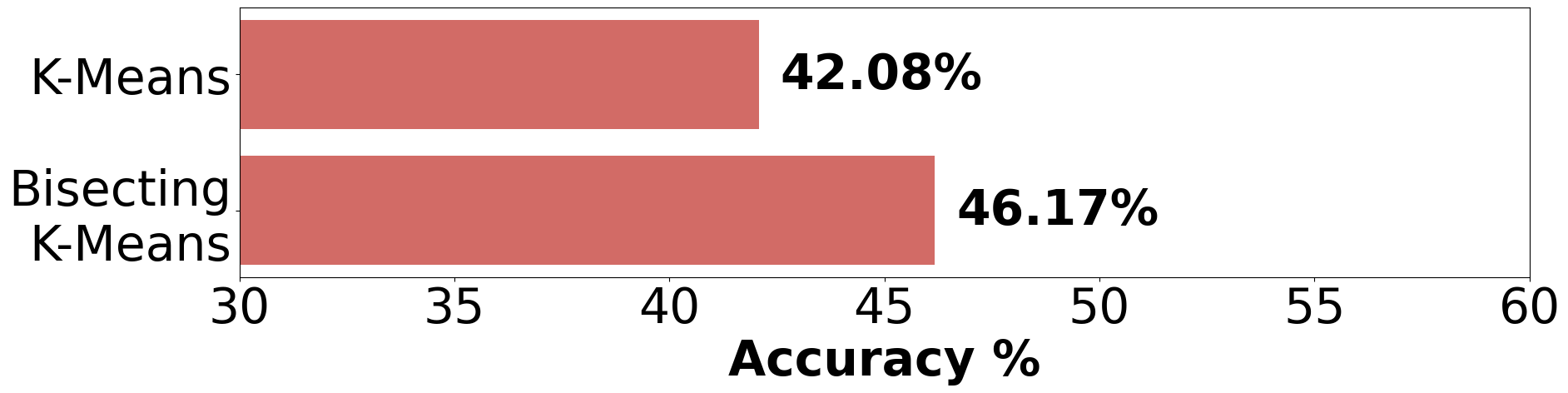}
    \caption{Performance difference between regular K-Means and Bisecting K-Means clustering, all performed using the phi-2 model.}
    \label{fig:K-Means-methods-performance}
\end{figure}

To improve the effectiveness of RAG, it is essential to group similar data, making it easier to identify patterns and trends within telecommunications data. To do this, the Bisecting K-Means clustering method is employed on the embedding vectors calculated in each chunk. This method is a variant of the K-Means algorithm that aims to improve the quality of the clusters formed, offering advantages in the segmentation of large volumes of data\cite{7847231}. Unlike the traditional technique, which partitions the data into $K$ clusters simultaneously, Bisecting K-Means adopts a hierarchical divisive approach, dividing the data into smaller subgroups iteratively. Figure \ref{fig:K-Means-methods-performance} represents the difference in accuracy that justifies using Bisecting K-Means instead of the traditional method.

This process can be described in the following steps: at the time of initialization of the algorithm, all documents are kept in a single cluster. Then, the cluster is divided into two clusters using the K-Means algorithm, i.e., $B = 2$. Then, all the calculated clusters are checked in search of the cluster with the most considerable Sum of Square Error (SSE). Then, the cluster with the highest SSE is recursively split into two more clusters with the K-Means algorithm with $B = 2$, and this process is repeated until the desired cluster values are generated. 

The SSE was used to measure the variability within a cluster by calculating the sum of the squares of the distances between each data point and the cluster centroid. The lower the SSE, the more homogeneous the data within the cluster, indicating a better quality in the data segmentation. Using the SSE allows us to identify which clusters have more significant internal variability and, therefore, which should be subdivided to improve the accuracy of the clustering. SSE is calculated as follows: 

\begin{equation}
\label{eq:sse}
SSE = \sum_{i=0}^{n}(X_{i}-\bar{X})^{2}
\end{equation}
where $X_{i}$ represents each data point, and $\bar{X}$ represents the mean of the points in the cluster. This work tested several cluster values to determine the ideal configuration.

In this work the embeddings were grouped into 18 clusters, grouping those with similar values. This simplified the identification of groups of data with similar characteristics and facilitated the recovery of relevant information during text generation.

\subsection{Routing Module}

The RAG routing module enhances the ability to generate more accurate and contextually relevant responses by retrieving pertinent information from a vast repository of documents. This module is necessary because it bridges the gap between static knowledge bases and dynamic, query-specific information needs, thereby improving the quality and relevance of generated content. 

\begin{figure}[ht]
    \centering
    \includegraphics[width=1\linewidth]{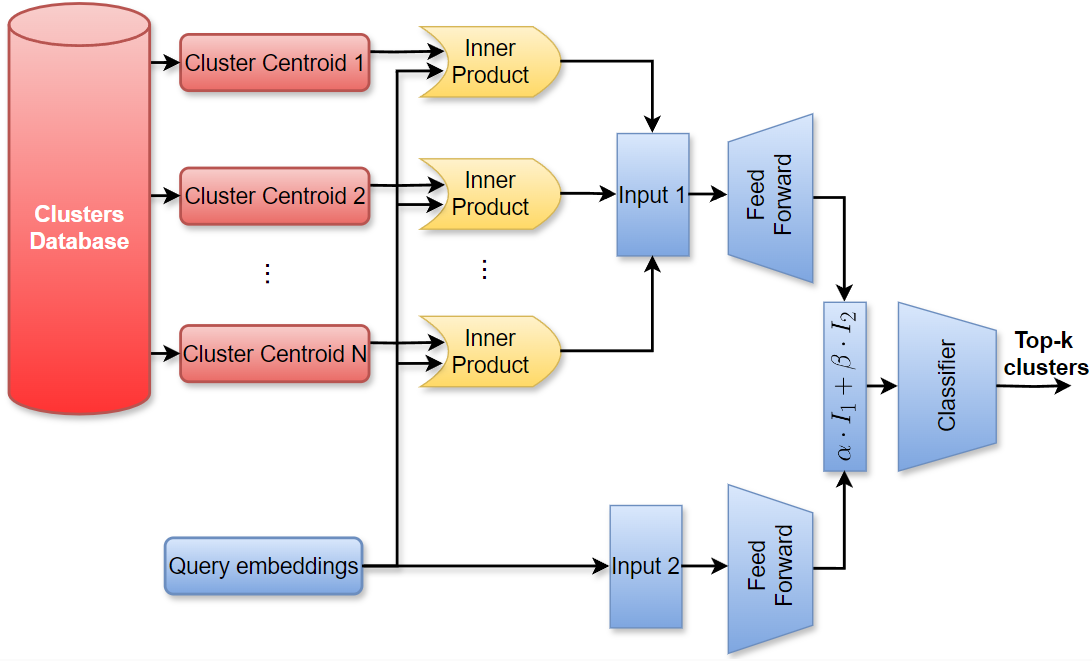}
    \caption{Routing module.}
    \label{fig:Router_diagram}
\end{figure}

As shown in Figure \ref{fig:Router_diagram}, the routing module takes as input $\mathcal{D}$ and $\textbf{Q}$. From the former, the module obtains the centroid of each cluster from the vector database - which can be represented as embedding vectors - and these vectors are compared to the embedded user entry embeddings $\textbf{Q}$ using inner products. This comparison produces a vector with a number of elements equal to the number of clusters 
$|\mathcal{D}|$ created during the preprocessing block. To refine this information, the vector undergoes an enhancement process with a Feed Forward Layer, which adjusts its size from $|\mathcal{D}|$ to 256 dimensions, forming $I_1$, which is referred to as Input 1. This number was selected due to being an intermediate number between the number of clusters and the dimension of $\textbf{Q}$.

Simultaneously, $\textbf{Q}$, typically of size 1024 due to the embedding model used (BAAI), is processed through another Feed Forward Layer to reduce its dimensionality to 256, forming $I_2$, referred to in Figure \ref{fig:Router_diagram} as Input 2. The two inputs, now both with 256 dimensions of size, are linearly combined to create the classifier input $I_c$, which can be defined as:

\begin{equation}
    I_c = \alpha I_1 + \beta I_2 
\end{equation}
where $\alpha$ and $\beta$ are constants. Then, $I_c$ is fed into a classifier layer, which is a softmax function that generates a probability distribution indicating which clusters are more related to the user's query. This distribution allows the system to interpret the relevance of each cluster to the query.

To optimize the retrieval process, the probability distribution is truncated to focus only on a subset of the clusters $\mathcal{K}$, known as the top-K clusters. In this work, this number was 8. By concentrating on this subset, the system can efficiently perform searches in the embedding space, narrowing down the relevant chunks of information without needing to search the entire space. This approach significantly enhances the system's efficiency and accuracy in retrieving the most pertinent information in response to the user's query. At the same time, it also decreases the time to perform searches in the vector database, as shown in Section IV.

\subsection{Fine-tuned Model}
 A fine-tuning process\footnote{\url{https://huggingface.co/SuLLMerica}} was employed to enhance a model's understanding of context and tailor its responses to meet the desired results. Based on the probable on-site usage, a small model was preferable. Therefore, phi-2 was selected because of its size, which allows us to fine-tune it and deploy the model on small devices.

To create the training dataset, we generated a context for each training example using RAG. By integrating the context in the training process, it was ensured that the model could better utilize relevant external information, thereby improving its ability to comprehend and respond accurately to domain-specific queries, as shown in\cite{ragfinetune}. Specifically, the context was concatenated with the terms, definitions, question, options, answer, and explanation of the answer, using the base chat template from Hugging Face to make it easier for the model to separate context, query, and response, as shown in Figure \ref{fig:train_example}.

\begin{figure}[ht]
\centering
\begin{tikzpicture}[
  every node/.style={font={\normalsize}},
  box/.style={rectangle, draw, rounded corners, align=left, text width=8cm, inner sep=3pt},
  arrow/.style={-{Latex[length=3mm]}, line width=1pt}
]
\node[box, fill=white] (context) {
<|im\_start|>context\\
(Retrieved context)\\
Terms and Definitions: (Terms and Definitions)\\
Abbreviations: (Abbreviations)\\
<|im\_end|>\\
<|im\_start|>user\\
Please provide the answer to the following multiple-choice question: (question and options)\\
Choose the correct option from the above options.\\
<|im\_end|>\\
<|im\_start|>assistant\\
The correct option number is option (correct option number): (option text)\\
Explanation: (explanation)\\
<|im\_end|>
};
\end{tikzpicture}
\caption{Training text example.}
\label{fig:train_example}
\end{figure}

A separate dataset was also created in which the answer referenced the specific section in the 3GPP documents from which the answer was derived. If the answer could not be found in the context, it explicitly states it wasn't found. This dataset was used to fine-tune a more reliable model because, with the citation, it would be possible to look up the section in the documents and confirm the answer. To generate this answer, the context, terms, abbreviations, question, options, answer, and explanation were fed to GPT-4 \cite{openai2024gpt4technicalreport} to extract the most valuable retrievals and documents and cite them in the answer. That dataset was only used to fine-tune the phi-2 model and was named Enhanced responses as in Figure \ref{fig:Stacked_graph}.

%Falar sobre arquitectura utilizada

% Falar sobre as métricas utilizadas

%Falar sobre os parametros do Finetuning, modelo de embedding

\section{Results}

This section presents the framework's performance in enhancing the capabilities of LLMs applied to the 3GPP Domain. All fine-tuning experiments were conducted using Hugging Face AutoTrain on a machine equipped with 8 vCPUs, 30 GB of RAM, and a single NVIDIA L4 GPU with 24 GB of VRAM.

\begin{figure}[ht]
    \centering
    \includegraphics[width=1\linewidth]{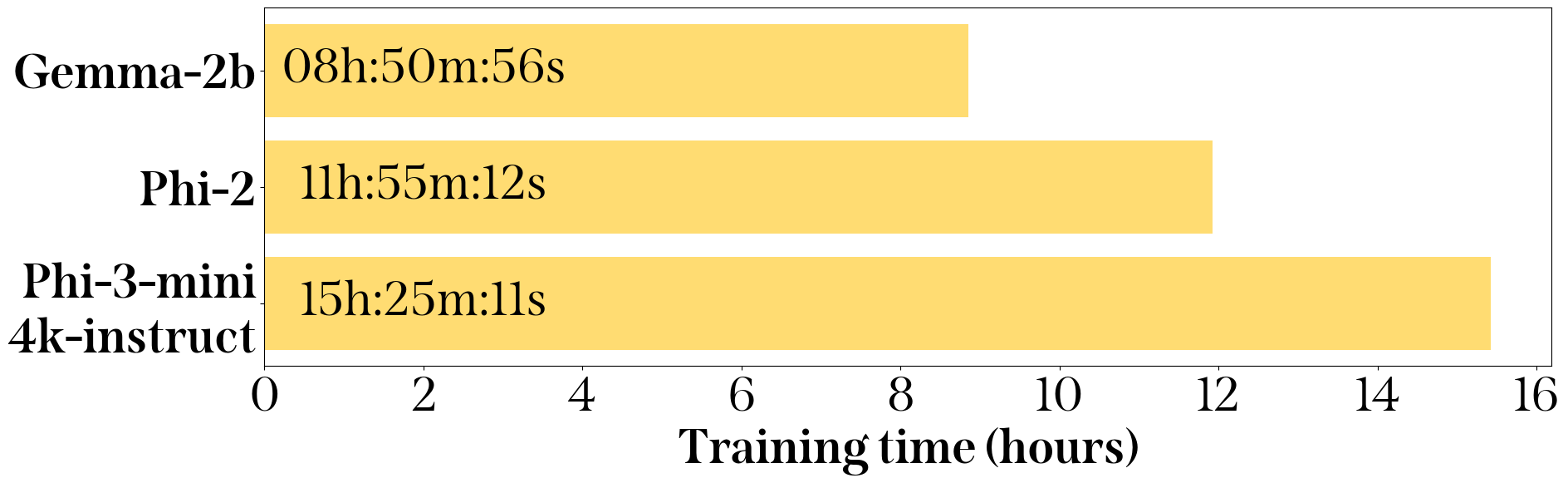}
    \caption{Time it took to fine-tune the models on 8vCPUs 30GB RAM and 1xL4 GPU 24GB VRAM in Hugging Face AutoTrain.}
    \label{fig:training_time}
\end{figure}

The research focused on the phi-2 model\footnote{\url{https://github.com/SuLLMerica/SuLLMerica}} due to its smaller size and shorter training time compared to Phi-3-4k-instruct \cite{phi3}, as illustrated in Figure \ref{fig:training_time}, as well as its better performance, in early testings with RAG, than gemma-2-2b\cite{gemma2} despite the similar size. After the development, those two other models were tested with the whole process, separating the gains to demonstrate the effect of each part of the system.

\begin{figure}[ht]
    \centering
    \includegraphics[width=1\linewidth]{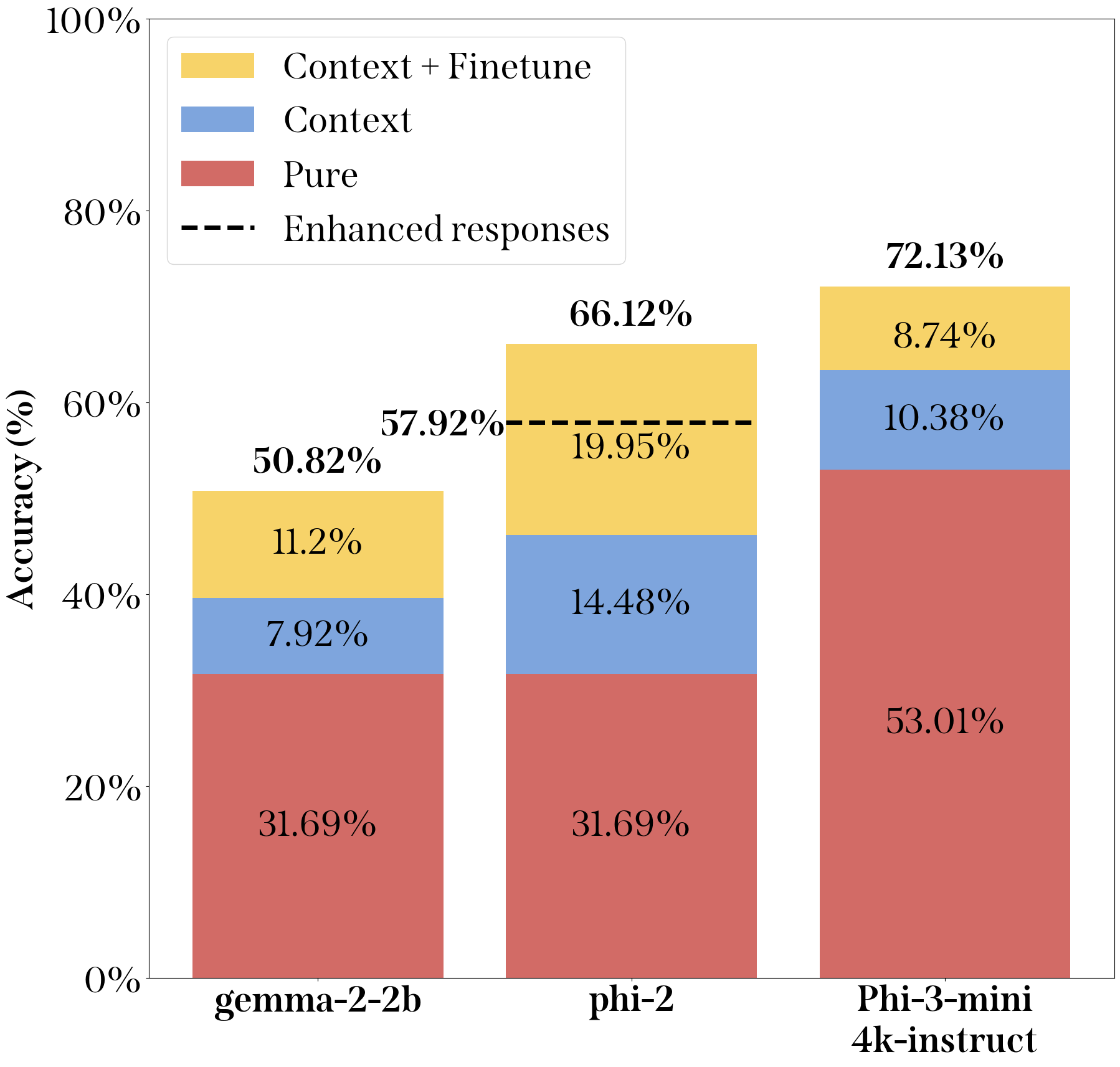}
    \caption{Performance improvement of each model separated by methods.}
    \label{fig:Stacked_graph}
\end{figure}

As shown in Figure \ref{fig:Stacked_graph}, substantial accuracy gains were observed by employing the described methods, getting up to a 34.4\% increase over the base phi-2 model, combining the RAG with fine-tuning. Furthermore, it was found that the technique also made the gemma-2-2b and Phi-3-4k-instruct model's accuracy increase substantially, indicating that this method's gains might be generalizable to other models and even bigger ones like Phi-3-mini-4k-instruct.

Additionally, it was found that the model trained with the enhanced response performed worse than the non-enhanced one, as shown in Figure \ref{fig:Stacked_graph}. Despite that, this model could be helpful because of its ability to relate the question to the 3GPP documents and provide a better starting point for finding the answer.

% Mostrar os resultados

\section{Future Works}

Although the proposed framework demonstrates advances in question-answering tasks, there are still many opportunities for performance enhancements. Future research could explore novel knowledge representation techniques, such as knowledge graphs, to improve information retrieval and reasoning capabilities. Additionally, greater emphasis on data preprocessing, particularly cleaning document chunks before converting them to embeddings, could significantly enhance the quality and relevance of the input data. Also, refining the routing module to output a more focused selection of clusters would likely lead to more efficient and accurate responses. Finally, using larger language models such as GPT-4 could yield substantial improvements, provided that careful evaluation is conducted to strike an optimal balance between model size, computational resources, and performance gains.

\section{Conclusions}

This work presented the integration of a routing module within the RAG architecture, combined with clustering techniques, significantly enhancing the efficiency and accuracy of information retrieval. By preprocessing 3GPP source documents through chunking, embedding, and Bisecting K-Means clustering, the system effectively organizes vast amounts of data into meaningful clusters. These clusters are stored in a vector database, providing a structured foundation that allows the RAG model to quickly access and utilize the most relevant information in response to user queries.

Ultimately, this approach to RAG architecture leverages the power of both the preprocessing phase, by gathering embeddings by content instead of using formats defined in the source documents, and the real-time processing phase to deliver highly accurate and contextually appropriate responses in achievable time. The experiment results show that using smaller LLMs, it's possible to obtain 66.12\% and 72.13\% accuracy in telecom documents using phi-2 and phi-3, respectively, while taking less than 12 and 16 hours of training time for both cases, respectively.

\section*{Acknowledgment}

This work was carried out with the guidance, mentorship, and financial investment of Dr. Aldebaro Klautau from LASSE — Research and Development Center in Telecommunications, Automation, and Electronics — and prof. Carlos Alberto Astudillo Trujillo from the Computer Networks Laboratory (LRC) at the State University of Campinas (Unicamp). We want to express our profound gratitude for their support in developing this research.

\bibliographystyle{ieeetr}
\bibliography{REFERENCES}

\end{document}